\documentclass[12pt,preprint]{aastex}
\usepackage{emulateapj5,epsf}

\shorttitle{Why small BH in disks?}
\shortauthors{Kawakatu et al.}

\begin{document}

\title{Why Massive Black Holes are Small in Disk Galaxies ?}


\author{Nozomu Kawakatu\altaffilmark{1} and Masayuki Umemura\altaffilmark{2}}
\affil{Center for Computational Physics, University of
  Tsukuba, Ten-nodai, 1-1-1 Tsukuba, Ibaraki, 305-8577, Japan}

\altaffiltext{1}{kawakatu@rccp.tsukuba.ac.jp}
\altaffiltext{2}{umemura@rccp.tsukuba.ac.jp}



\begin{abstract}
A potential mechanism is proposed to account for the fact that
supermassive black holes (SMBHs) in disk galaxies appear to be smaller than 
those in elliptical galaxies in the same luminosity range.
We consider the formation of SMBHs by radiation drag 
(Poynting-Robertson effect), which extracts angular momentum 
from interstellar matter and thereby 
drives the mass accretion onto a galactic center. 
In particular, we quantitatively scrutinize the efficiency of radiation drag 
in a galaxy composed of bulge and disk, and elucidate the relation 
between the final mass of SMBH and the bulge-to-disk ratio of the galaxy. 
As a result, it is found that the BH-to-galaxy mass ratio,
$M_{\rm BH}/M_{\rm galaxy}$, decreases with a smaller bulge-to-disk ratio,
due to the effects of geometrical dilution and opacity, and is reduced 
maximally by two orders of magnitude, resulting in
$M_{\rm BH}/M_{\rm galaxy}\approx 10^{-5}$.
In contrast, if only the bulge components in galaxies are focused,
the BH-to-bulge mass ratio becomes $M_{\rm BH}/M_{\rm bulge}\approx 10^{-3}$, 
which is similar to that found in elliptical galaxies. 
Thus, it turns out that the mass of SMBH primarily correlates with
a bulge, not with a disk, consistently with observational data.
\end{abstract}

\section{Introduction}
\label{INTRO}
Supermassive black holes (SMBHs) are now thought to reside in virtually 
all galactic bulges.
The recent compilation of the kinematical data of galactic centers 
in both active and inactive galaxies has shown that 
the mass of SMBH does correlate with the mass of a galactic bulge; 
the mass ratio of BH to bulge is $\approx 0.001$ as a median value 
(Kormendy \& Richstone 1995; 
Richstone et al. 1998; Magorrian et al. 1998; Gebhardt et al. 2000; Ferrarese \& 
Merritt 2000; Merritt \& Ferrarese 2001; McLure \& Dunlop 2001, 2002; 
Wandel 2002).
On the other hand, it appears that the SMBH does not correlate with a disk component (Kormendy \& Gebhardt 2001).
In disk galaxies, the mass ratio is significantly smaller than 
0.001 if disk stars are included (Salucci et al.2000; Sarzi et al. 2001, Ferrarese 2002, Baes et al. 2003). 
These observations imply that the formation of the SMBH is related to the bulge components of galaxies.

Recently, as a possible mechanism for the formation of SMBHs, 
the radiation-hydrodynamic model is
proposed, where the radiation drag (Poynting-Robertson effect) 
due to stellar radiation in a galactic bulge
drives the mass accretion to grow a massive BH (Umemura 2001).
This is a parallel model of the formation of massive BHs by
the Compton drag in an early universe
(Umemura, Loeb, \& Turner 1997).
In an optically thick regime, the efficiency of radiation drag is saturated 
due to the conservation of the photon number (Tsuribe \& Umemura 1997).
The timescale of radiation drag-induced mass accretion is $t_{\rm drag}=
8.6\times 10^{7}R^{2}_{\rm kpc}
(L_{*}/10^{12}L_{\odot})^{-1}(Z/Z_{\odot})^{-1}{\rm yr}$, 
where $R_{\rm kpc}$ is the galaxy size in units of kpc, and $Z$ is the 
metallicity of gas.
The mass accretion rate is given by 
$\dot{M}=-M_{\rm g}d \ln J/dt \simeq L_{*}/c^{2}$,
where $J$, $L_{*}$, and $M_{\rm g}$ are 
the total angular momentum of gaseous component, 
the total luminosity of the system, and the total mass of gas, respectively.
Thus, the final BH mass is estimated by 
$M_{\rm BH} \simeq \int L_{*}(t)/c^{2}dt.$
Hence, the final BH mass is proportional to the total radiation energy from stars,
 and the theoretical upper limit of BH-to-bulge mass ratio is determined by the energy 
conversion efficiency of nuclear fusion from hydrogen to helium, i.e., 0.007 (Umemura 2001).
In practice, the inhomogeneity of interstellar 
medium helps the radiation drag to sustain high efficiency
(Kawakatu \& Umemura 2002).
By incorporating the realistic chemical evolution, the radiation 
drag model predicts $M_{\rm BH}/M_{\rm bulge}\simeq 0.001$ 
(Kawakatu, Umemura, \& Mori 2003).
However, the radiation drag efficiency could be strongly subject to 
the effect of geometry (Umemura et al. 1997,1998; Ohsuga et al. 1999a).
If the system is spherical, the emitted photons are effectively 
consumed within the system, 
whereas a large fraction of photons can escape from disk-like systems, and thus 
the drag efficiency is likely to be reduced considerably.
It is suggested that such a geometrical effect may be the reason 
why the BH is smaller in disk galaxies (Umemura 2001; Umemura 2003),
although the details have not been clear quantitatively.

In this Letter, we scrutinize the efficiency of radiation drag in a system
composed of bulge and disk.
For the purpose, we accurately solve the 3D radiative transfer 
in an inhomogeneous interstellar medium.
The main goal is to elucidate the relationship between the final BH mass 
formed by the radiation drag and the bulge-to-disk ratio of host galaxy.

\section{Model}
We consider a system composed of bulge and disk, which consist of 
dark matter, stars, and dusty interstellar matter (ISM).
The total baryonic mass of the system, $M_{\rm b}$,
is set to be $10^{11}M_{\odot}$.
But, it is noted that the results are scaled with $M_{\rm b}$.
The dark matter is distributed following the profile by 
Navarro, Frenk, $\&$ White (1997),
$\rho (r) \propto (r/r_{\rm s})^{-1}(1+r/r_{\rm s})^{-2}$,
where $r_{\rm s}$ is the characteristic radius, which is
$r_{\rm s}=50{\rm kpc}$ here.
The mass of the dark matter component, $M_{\rm DM}$, 
is equal to $M_{\rm b}$ within the galaxy size of $r_{\rm g}=$10kpc.

We incorporate the evolution of galaxy from the formation stage,
where a galactic system starts with a protogalactic cloud purely 
composed of barynic gas. By invoking the star formation in Schmidt's law, 
stellar components are generated in this system.
To include the recycling of ISM, 
we employ an evolutionary spectral synthesis 
code, PEGASE (Fioc \& Rocca-Volmerange 1997), 
with assuming a Salpeter-type initial mass function
 (IMF) as $\phi = dn/d\log m_{*}=\phi_0 (m_{*}/M_{\odot})^{-1.35}$ 
for the mass range of $[0.1 M_\odot,~60M_\odot]$, 
where $m_{*}$ is the stellar mass.
The star formation rate (SFR) per unit mass at time $t$, $C(t)$, 
is assumed to be proportional to the gas mass fraction, 
$C(t) = kf_{\rm g}$ with $f_{\rm g}\equiv M_{\rm g}(t)/M_{\rm b}$,
where $k$ is a rate coefficient which is $8.6 {\rm Gyr}^{-1}$ 
for the bulge component and $0.86 {\rm Gyr}^{-1}$ 
for the disk component, respectively. 

The formed stars in the bulge
are distributed following Hernquist's profile (1990), 
which mimics de Vaucouleur's profile: 
$\rho (r) \propto (r/r_{\rm core})^{-1}(1+r/r_{\rm core})^{-3}$ ,
where $r_{\rm core}$ is the core radius.
The bulge radius, $r_{\rm bulge}$, is given by a mass-to-size relation as
$ r_{\rm bulge}/r_{\rm g}=(M_{\rm bulge}/M_{\rm galaxy})^{1/2}$
(Kormendy 1985; Chiosi \& Carraro 2002),
where $ M_{\rm bulge}$ is the bulge mass and $ M_{\rm galaxy}$ 
is the final stellar mass of the galaxy, and 
$r_{\rm core}$ is set to $r_{\rm core}=0.1r_{\rm bulge}$.
In the present analysis, the bulge mass fraction 
($M_{\rm bulge}/M_{\rm galaxy}$) is altered as a key parameter.
In addition, angular momentum is smeared, 
assuming rigid rotation, according to the spin parameter of
$\lambda=0.05$ (Barnes \& Efstathiou 1987; Heavens \& Peacock 1988).

As for the disk component, formed stars are distributed exponentially 
in the radial direction, with assuming
the vertical scale-height which is changed from $0.01r_{\rm g}$
to $0.1r_{\rm g}$.
The disk radius, $r_{\rm disk}$, is given by a mass-to-size relation of 
$ r_{\rm disk}/r_{\rm g}=(M_{\rm disk}/M_{\rm galaxy})^{1/2}$,
where $M_{\rm disk}$ is the disk mass.
The disk component is assumed to be in rotation balance.

As for ISM, we consider clumpy matter, since the ISM is observed to be 
highly inhomogeneous in active star-forming galaxies 
(Sanders et al. 1988; Gordon, Calzetti, \& Witt 1997). 
Here, $N_{\rm c}(=10^{4})$ identical clouds are distributed randomly 
in a system of the bulge and disk. 
The number of clouds is minimal enough to resolve the 
disk scal-height of $0.01r_{\rm g}$. (It is noted that simulations 
with a three times larger number of clouds did not lead 
to any fundamental difference.)
The distribution and motion 
of the ISM is the same as those of stellar components.
We assume that the cloud covering factor is unity
according to the analysis by Kawakatu \& Umemura (2002).
The internal density in a cloud is assumed to be uniform,
and the opacity is determined by a dust-to-gas ratio 
which is calculated by PEGASE.
The optical depth of a cloud and therefore 
the overall optical depth of galaxy, $\tau_{\rm T}$, depend on the cloud size
$r_{\rm c}$, where $r_{\rm c}=100$pc is assumed as a fiducial case.
But we have confirmed that no essential difference in the final BH mass is 
found by changing $r_{\rm c}$ so as to enhance $\tau_{\rm T}$ by an order. 

\vspace{5mm}
\epsfxsize=8cm 
\epsfbox{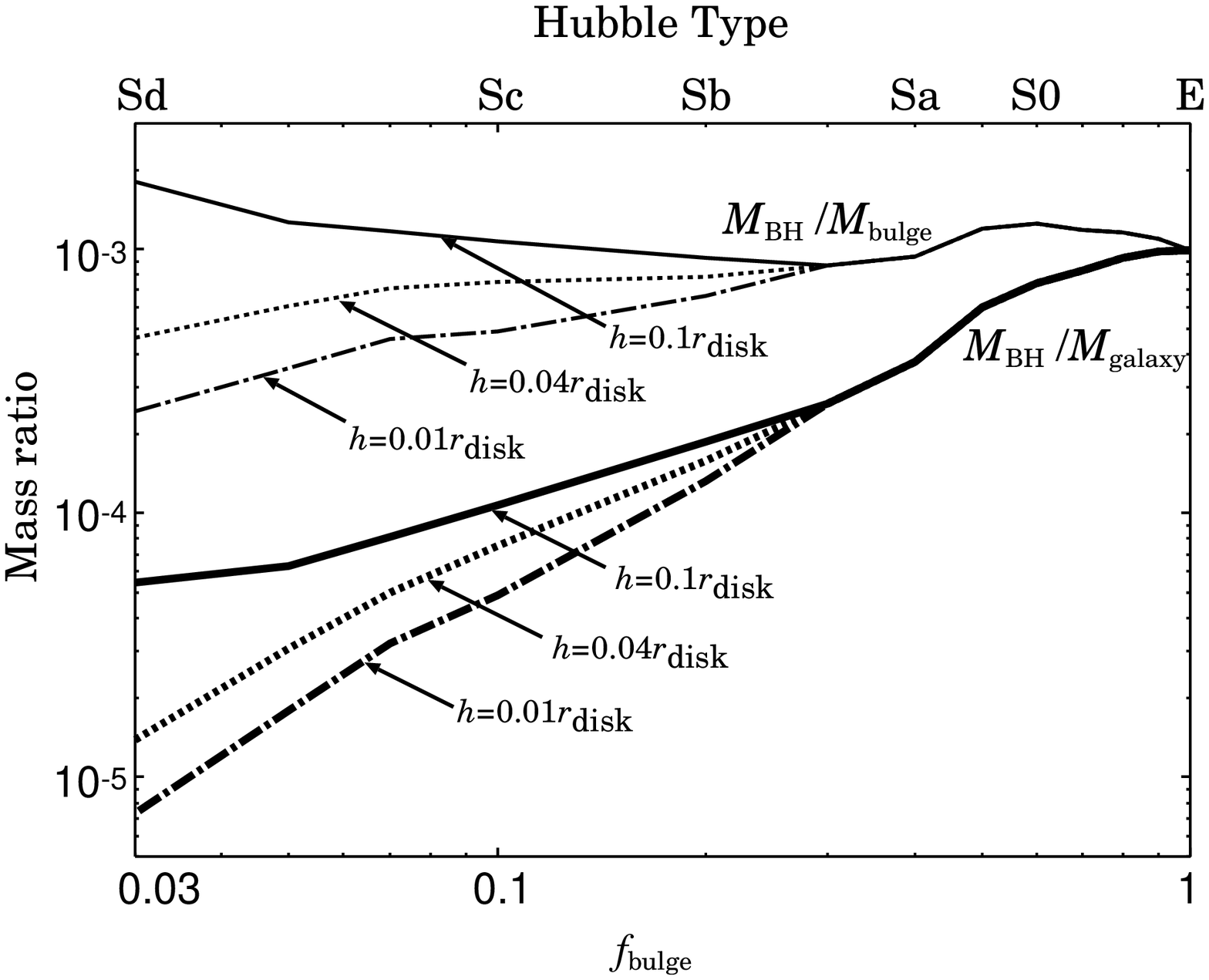}
\figcaption{
The BH-to-galaxy mass ratio 
($M_{\rm BH}/M_{\rm galaxy}$) and
the BH-to-bulge mass ratio 
($M_{\rm BH}/M_{\rm bulge}$) are shown against 
the bulge fraction, $f_{\rm bulge}\equiv M_{\rm bulge}/M_{\rm galaxy}$.
The thick lines show $M_{\rm BH}/M_{\rm galaxy}$ with changing
the disk scale-height from $h=0.01r_{\rm disk}$ up to 
$h=0.1r_{\rm disk}$ and
the thin lines do $M_{\rm BH}/M_{\rm bulge}$.
The top abscissa is the indication for 
the Hubble type, which is determined by the mean bulge-to-disk 
ratio for disk galaxies (Kent 1985).
\label{Figure 1}
}
\vspace{5mm}

The radiation drag, which drives the mass accretion, originates 
in the relativistic effect in absorption and reemission of radiation.
This effect is involved in relativistic radiation hydrodynamics (RHD) 
(Umemura, Fukue, \& Mineshige 1997, 1998; Fukue, Umemura, \& Mineshige 1997).
The angular momentum transfer in RHD 
is given by the azimuthal equation of motion in cylindrical coordinates,
\begin{equation}
\frac{1}{r}\frac{d(rv_{\rm \phi})}{dt}=\frac{\chi}{c}[F^{\rm \phi}-
(E+P^{\rm \phi \phi})v_{\rm \phi}], \label{ldot}
\end{equation} 
where $\chi$ is the mass extinction coefficient,
$E$ is the radiation energy density, $F^{\rm \phi}$ 
is the radiation flux, and $P^{\rm \phi\phi}$ is the radiation stress tensor.
All radiative quantities are determined by radiation from stars diluted by dusty ISM.
Here, we solve radiative transfer including the dust extinction by ISM and
obtain these radiative quantities. Then,
we evaluate the total angular momentum loss rate
(see Kawakatu \& Umemura 2002 for the details of method).
If an optically thick cloud is irradiated by the radiation,
an optically-thin surface layer is stripped by the radiation 
drag (Tsuribe \& Umemura 1997), and loses angular momentum.
Since the bulge luminosity is lower than 
the Eddington luminosity for dusty gas, $L_{Edd}=4\pi cGM_{\rm dyn}/\chi$
with the dynamical mass $M_{\rm dyn}$ within $r_{\rm g}$, 
during the whole history of galaxy, 
the stripped gas is not blown away by the radiation pressure, but confined 
in the galaxy.
Hence, the gas moves in a gravitational potential weakened
by radiation pressure and loses angular momentum by radiation drag.
The stripped gas losing angular momentum 
falls onto the galactic center to form a massive object, 
which is likely to undergo the inside-out viscous 
collapse (Mineshige, Tsuribe, \& Umemura 1998; Tsuribe 1999) 
and eventually evolve into a SMBH by the general relativistic instability 
(Saijo et al. 2002; Shibata \& Shapiro 2002).
Using equation (\ref{ldot}), we can calculate the
mass accretion rate onto the central object by
$\dot{M}_{\rm g}/M_{\rm g}=-\dot{J}/J$, where $J$ is the total 
angular momentum of ISM. Then,  
the BH mass ($M_{\rm BH}$) is assessed as     
\begin{equation}
M_{\rm BH}=\int_{0}^{t_0}\dot{M}_{\rm g}dt 
=-\int_{0}^{t_0}M_{\rm g}\frac{\dot{J}}{J}dt,
\end{equation}
where $t_{0}$ is the Hubble time.

\vspace{5mm}
\epsfxsize=8cm 
\epsfbox{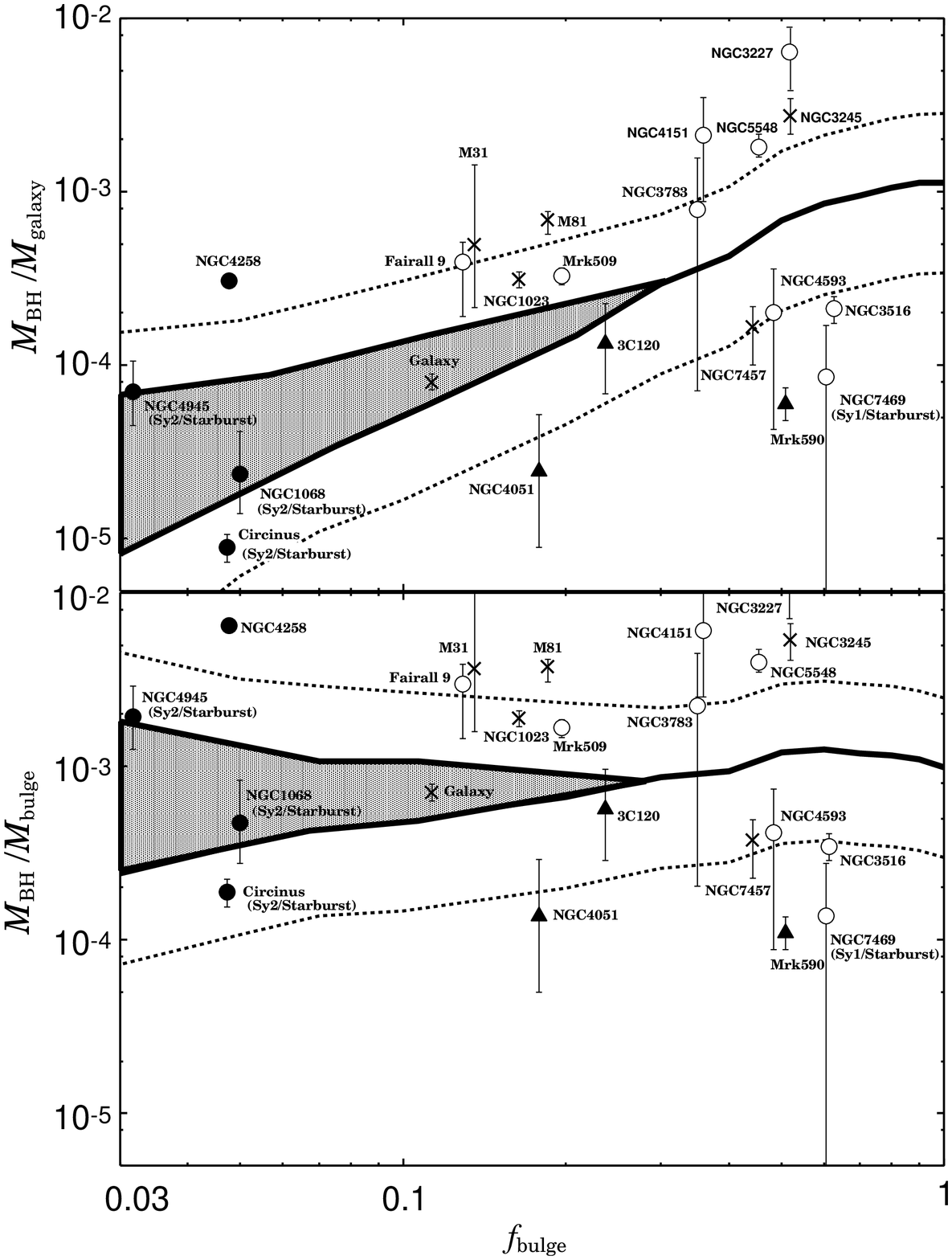}
\figcaption{
Comparison between the theoretical prediction and the observational data
on BH mass for disk galaxies.
The upper panel shows the BH-to-galaxy mass ratio 
($M_{\rm BH}/M_{\rm galaxy}$) and
the lower panel shows the BH-to-bulge mass ratio 
($M_{\rm BH}/M_{\rm bulge}$) against 
the bulge fraction, $f_{\rm bulge}\equiv M_{\rm bulge}/M_{\rm galaxy}$.
The observational data are based upon Table 1 and plotted
by symbols: 
{\it Crosses} -- disk galaxies which do not possess AGNs,
{\it open circles} -- Seyfert 1 galaxies (Sy1s),
{\it filled triangles} -- narrow line Seyfert 1 galaxies (NLSy1s),
and {\it filled circles} -- Seyfert 2 galaxies (Sy2s).
Seyfert galaxies accompanied by starburst activities are specified like
Sy1/starburst or Sy2/starburst.
The hatched area is the prediction shown in Fig. 1. 
Two dotted lines in each panel are the upper and lower bound of 
the theoretical prediction by including uncertainties 
other than the geometrical effect (see the end of \S3).
\label{Figure 2}
}

\section{Results}
By changing the bulge mass fraction, 
$ f_{\rm bulge}\equiv M_{\rm bulge}/M_{\rm galaxy}$,
we investigate the relation between the final BH mass and 
the bulge-to-disk ratio of galaxy.
The growth of BH depends upon the optical depth 
of galaxy, $\tau_{\rm T}$, but no essential difference 
in the final BH mass is found by enhancing $\tau_{\rm T}$ by an order, 
because the mass accretion rate is saturated when the system 
becomes optically-thick
(Umemura 2001; Kawakatu \& Umemura 2002).
In Figure 1, the resultant BH-to-galaxy mass ratio,
$M_{\rm BH}/M_{\rm galaxy}$, is shown against
the bulge mass fraction ($f_{\rm bulge}$)
by thick lines for three values of disk scale-height.
We can see a clear trend that $M_{\rm BH}/M_{\rm galaxy}$ decreases
with decreasing $f_{\rm bulge}$.
This tendency comes from three effects:
First, a larger number of photons escape from the disk surface,
since the surface-to-volume ratio is larger for the disk component
compared to the bulge.
This effect is called the {\em geometrical dilution} here.
Secondly, the radiation from disk stars is heavily diminished across 
the disk, because the edge-on optical depth becomes larger quickly.
Thirdly, the velocity difference between a star and an absorbing cloud 
becomes closer to zero in an optically-thick disk, 
so that the radiation drag cannot work effectively.
For instance, if $f_{\rm bulge}=0.03$
which corresponds to Sd galaxies,
$M_{\rm BH}/M_{\rm galaxy}$ is reduced by a factor of $\approx 20$
to $\approx 100$, compared to $M_{\rm BH}/M_{\rm galaxy}\approx 0.001$ 
for elliptical galaxies ($f_{\rm bulge}=1$).
Furthermore, it is predicted that a bulgeless (pure disk) system 
can also possess a massive black hole with the mass ratio of
$M_{\rm BH}/M_{\rm galaxy}\simeq 0.001\times \left(h/2r_{\rm disk}\right)$, 
where $h$ is the scale height of disk.
This ranges from $6\times 10^{-6}$ to $5\times 10^{-5}$, 
corresponding to $h=0.01r_{\rm disk}$ and $h=0.1r_{\rm disk}$, respectively.

On the other hand, 
the bulge component itself is likely to be impervious to 
the above three geometrical effects. 
In Figure 1, by focusing on only the bulge component,
the BH-to-bulge mass ratio, $M_{\rm BH}/M_{\rm bulge}$,
is also shown. As seen in Figure 1, $M_{\rm BH}/M_{\rm bulge}$ is
almost constant as expected.
Hence, it is concluded that the BH-to-galaxy mass ratio 
in disk galaxies is substantially determined by 
bulge components, not by disk components:
$M_{\rm BH}/M_{\rm galaxy}\simeq 0.001$,
where the constant value of 0.001 is basically determined 
by the energy conversion efficiency of nuclear fusion from 
hydrogen to helium, i.e., 0.007 (Umemura 2001).  
The BH accretion can trigger an active galactic nucleus (AGN) 
as well, and then the further mass accretion is
induced via radiation drag by the AGN.
The effect can enhance the final BH mass maximally by a factor of 1.7 
(Umemura 2001).
Also, a different level of ISM clumpiness can reduce the final BH mass 
by a factor of 2 (Kawakatu \& Umemura 2002). 
As for the effect of IMF, if the slope and the mass range of IMF 
are changed to satisfy the spectrophotometric properties, 
then the final BH mass is 
altered by a factor of $\pm 40\%$ .
Similar uncertainties attach also to the final BH mass in disk galaxies.

\section{Comparison with Observations}
The present results are compared with observational data for SMBHs in disk
galaxies in Figure 2. The data we adopted are summarized in Table 1.
In the upper panel of Figure 2, $M_{\rm BH}/M_{\rm galaxy}$ is shown 
against $f_{\rm bulge}$.
The hatched area is the prediction shown in Figure 1. 
Two dotted lines are respectively the upper and lower bound of 
the prediction by including other theoretical uncertainties 
mentioned above.
The observational data are categorized into four types:
disk galaxies which do not possess AGNs,
Seyfert 1 galaxies (Sy1s),
narrow line Seyfert 1 galaxies (NLSy1s),
and Seyfert 2 galaxies (Sy2s).
Also, the objects accompanied with starburst events are specified.
The observational data in Figure 2
shows the decrease of $M_{\rm BH}/M_{\rm galaxy}$ by
more than an order of magnitude from $f_{\rm bulge}=1$ to $f_{\rm bulge}=0.03$.
This trend is broadly consistent with the theoretical prediction.
But, it is worth noting that several objects possess relatively small BHs
compared to the prediction, and they are mostly Seyfert galaxies
with starburst events (NGC 7469, NGC 1068, Circinus) 
or NLSy1s (Mrk 590, NGC4051).
In recent years, an evolutionary model for AGNs beyond the AGN
unified theory has been considered from both observational and theoretical
points of view (Maiolino et al. 1995; Radovich, Rafanelli, \& Barbon 1998; 
Ohsuga \& Umemura 1999, 2001). In this model, AGNs with starbursts
are in earlier evolutionary phase, and therefore the BH mass is likely
to be smaller, compared to a post-starburst phase. 
Furthermore, NLSy1s are thought to be in a rapidly growing phase 
of an immature BH (Mathur et al. 2001). NGC1068 could be also 
in such a phase (Kawaguchi 2003). This point may be intriguing to analyze
further with larger sample data in the future.

In the lower panel of Figure 2, $M_{\rm BH}/M_{\rm bulge}$ is shown 
against $f_{\rm bulge}$.
When only the bulge component is focused, the observational data 
for $M_{\rm BH}/M_{\rm bulge}$ lies at a level of 
$\approx 0.001$, and they roughly agree 
with the prediction. But, again the ratios for Seyferts with starbursts and 
NLSy1s fall appreciably below $0.001$.

The present model also predicts a SMBH even in a pure disk system
without bulge, and the mass of BH depends upon the disk scale-height. 
NGC4395 is an archetype of bulgeless galaxy in which a massive BH
resides (Filipenko \& Ho 2003). 
The BH fraction in this galaxy is estimated to be 
$M_{\rm BH}/M_{\rm galaxy}\simeq 1.5\times 10^{-5}$.
In the present model, this level can be realized by a disk scale height 
of $h \approx 0.04r_{\rm disk}$.

\acknowledgments
We thank T.Nakamoto, T.Saitou and A.Yonehara for helpful discussions.
We also thank the anonymous referee for precious comments.
Numerical simulations were performed with facilities at the Center of Computational Physics, 
University of Tsukuba. This work was supported in part by the Grant-in-Aid of the JSPS, 11640225.


\begin{table}[t]
\begin{center}
Table 1. SMBHs in Disk Galaxies \\[3mm]
{\scriptsize
\begin{tabular}{lcllrccc}
\hline \hline
Object & Type & $M_{\rm galaxy}(M_\odot)$ & $M_{\rm bulge}(M_\odot)$  & $M_{\rm BH}(M_\odot)~~~~~$  & AGN & Starburst & Reference \\
\hline
Galaxy  &  Sbc  &  $4.6\times 10^{10}$  &  $5.2\times 10^{9}$  & $3.7(\pm 0.4)\times 10^{6}$  &  ---  & --- & 1 \\
M31  &  Sb  &  $1.4\times 10^{11}$  &  $1.9\times 10^{10}$  & $7(+13-4)\times 10^{7}$  &  ---  & --- & 2 \\
M81  &  Sb  &  $9.7\times 10^{10}$  &  $1.8\times 10^{10}$  & $6.8(+0.7-1.3)\times 10^{7}$  &  ---  & --- & 3 \\
NGC1023  &  SB0  &  $1.4\times 10^{11}$  &  $2.3\times 10^{10}$  & $4.4(+0.4-0.5)\times 10^{7}$  &  ---  & --- & 4 \\
NGC3245  &  S0  &  $7.5\times 10^{10}$  &  $3.9\times 10^{10}$  & $2.1(\pm 0.5)\times 10^{8}$  &  ---  & --- & 5 \\
NGC7457  &  S0  &  $2.1\times 10^{10}$  &  $9.3\times 10^{9}$  & $3.5(+1.1-1.4)\times 10^{6}$  &  ---  & --- & 6 \\
Fairall 9  &  S  &  $2.2\times 10^{11}$  &  $2.9\times 10^{10}$  & $8.7(+2.6-4.5)\times 10^{7}$  &  Sy1  & --- & 7 \\
Mrk509  &  S  &  $2.9\times 10^{11}$  &  $5.7\times 10^{10}$  & $9.5(+1.1-1.1)\times 10^{7}$  &  Sy1  & --- & 7 \\
NGC3516  &  SB0  &  $8.0\times 10^{10}$  &  $4.9\times 10^{10}$  & $1.7(\pm 0.3)\times 10^7$  &  Sy1  & --- & 8 \\
NGC3783  &  SBa  &  $1.4\times 10^{10}$  &  $4.9\times 10^{9}$  & $1.1(+1.1-1.0)\times 10^7$  &  Sy1  & --- & 7 \\
NGC4151  &  SA  &  $5.7\times 10^{9}$  &  $2.0\times 10^{9}$  & $1.2(+0.8-0.7)\times 10^7$  &  Sy1  & --- & 7 \\
NGC3227  &  SAa &  $5.6\times 10^{9}$  &  $2.9\times 10^{9}$  & $3.6(\pm 1.4)\times 10^7 $  & Sy1 & --- & 8 \\
NGC4593  &  SBb  &  $3.3\times 10^{10}$  &  $1.6\times 10^{10}$  & $6.6(\pm 5.2)\times 10^6$  &  Sy1  & --- & 8 \\
NGC5548  &  Sa  &  $4.4\times 10^{10}$  &  $2.0\times 10^{10}$  & $8.0(+1.5-1.0)\times 10^7$  &  Sy1  & --- & 7 \\
NGC7469  &  SA  &  $8.9\times 10^{10}$  &  $5.5\times 10^{10}$  & $7.6(+7.5-7.6)\times 10^6$  &  Sy1  & Nuclear & 7 \\
3C120  &  S0  &  $2.2\times 10^{11}$  &  $5.2\times 10^{10}$  & $3.0(+2.0-1.5)\times 10^7$  &  NLSy1  & --- & 7 \\
Mrk590  &  Sa  &  $2.3\times 10^{11}$  &  $1.1\times 10^{11}$  & $1.4(+0.3-0.3)\times 10^7$  &  NLSy1  & --- & 7 \\
NGC4051  &  SB  &  $5.6\times 10^{10}$  &  $1.1\times 10^{10}$  & $1.4(+1.5-0.9)\times 10^6$  &  NLSy1  & --- & 7 \\
Circinus  &  Sb-Sd  &  $1.9\times 10^{11}$  &  $9.0\times 10^{0}$  & $1.7(\pm 0.3)\times 10^6$  &  Sy2  & Nuclear & 9 \\
NGC1068  &  Sb  &  $7.2\times 10^{11}$  &  $3.6\times 10^{10}$  & $1.7(+1.3-0.7)\times 10^7$  &  Sy2  & Circumnuclear & 10 \\
NGC4258  &  Sbc  &  $1.3\times 10^{11}$  &  $6.2\times 10^{9}$  & $4.0(\pm 0.1)\times 10^7$  &  Sy2  & --- & 11 \\
NGC4945  &  Scd  &  $2.4\times 10^{10}$  &  $7.2\times 10^{8}$  & $1.4(+0.7-0.5)\times 10^6$  &  Sy2  & Nuclear & 12 \\
\hline
\end{tabular}
}
\noindent
\end{center}
{\scriptsize References:
1) Sch$\ddot{\rm o}$del et al. 2003,
2) Bender et al. 2003,
3) Bower et al. 2000,
4) Bower et al. 2001,
5) Barth et al. 2001,
6) Gebhardt et al. 2003,
7) Wandel 1999,
8) Onken et al. 2003,
9) Greenhill et al. 2003,
10) Greenhill et al. 1996,
11) Miyoshi et al. 1995, 
12) Greenhill et al. 1997}
\end{table}


\end{document}